\documentclass[10pt,preprint]{aastex}

\usepackage{pdfpages}
\newcommand{\Rs}{$ R_{\odot}$}

\usepackage[small,compact,center]{titlesec}

  \renewenvironment{thebibliography}[1]{%
    \begin{oldthebibliography}{#1}%
      \setlength{\parskip}{0.3ex}%
      \setlength{\itemsep}{0.3ex}%
  }%
  {%
    \end{oldthebibliography}%
  }

\begin{document}

\section*{Exploring the Physics of the Corona with Total Solar Eclipse Observations}

\centerline{Shadia R. Habbal$^1$,  John Cooper$^2$, Adrian Daw$^3$, Adalbert Ding$^4$} 
\centerline{Miloslav Druckm\"uller$^5$, Ruth Esser$^6$, Judd Johnson$^7$ and Huw Morgan$^1$ }

\vskip 0.05in

\centerline{$^1$ \it Institute for Astronomy, University of Hawaii}
\centerline{$^2$ \it NASA/GSFC, Heliospheric Physics Laboratory, Code 672}
\centerline{$^3$ \it NASA/GSFC, Solar Physics Laboratory, Code 671}
\centerline{$^4$ \it Institute of Optics and Atomic Physics, Technische Universitaet Berlin, Germany}
\centerline{$^5$ \it Faculty of Mechanical Engineering, Brno University of Technology, Czech Republic}
\centerline{$^6$ \it University of Tromso, Norway}
\centerline{$^7$ \it Electricon, Boulder}

\section*{Abstract}
This white paper is a call for a concerted effort to support total solar eclipse observations over the next decade, in particular for the 21 August 2017 eclipse which will traverse the US continent. With the recent advances in image processing techniques and detector technology, the time is ripe to capitalize on the unique diagnostic tools available in the visible and near infrared wavelength range to explore the physics of the corona. The advantage of coronal emission lines in this wavelength range, over their extreme ultraviolet counterparts, is (1) the significant radiative component in their excitation process (in addition to the collisional excitation), which allows for observations out to a few solar radii, (2) the higher spectral selectivity available for imaging, giving well-defined temperature responses for each bandpass (one line as opposed to many), and (3) the capability of polarization measurements in a number of spectral lines. Consequently, the evolution of the thermodynamic and magnetic properties of the coronal plasma can be explored starting from the solar surface out to a few solar radii, namely the most important region of the corona where the expansion of the solar magnetic field and the acceleration of the solar wind occur. Since the planning of eclipse observations will not be possible without the invaluable NASA-published total solar eclipse bulletins by Espenak and Andersen, a call is also made to ensure continued support for these efforts.

\section*{Impetus: Examples from Recent Eclipse Observations}

\begin{figure}[ht]
\centerline{
\includegraphics[scale=0.5]{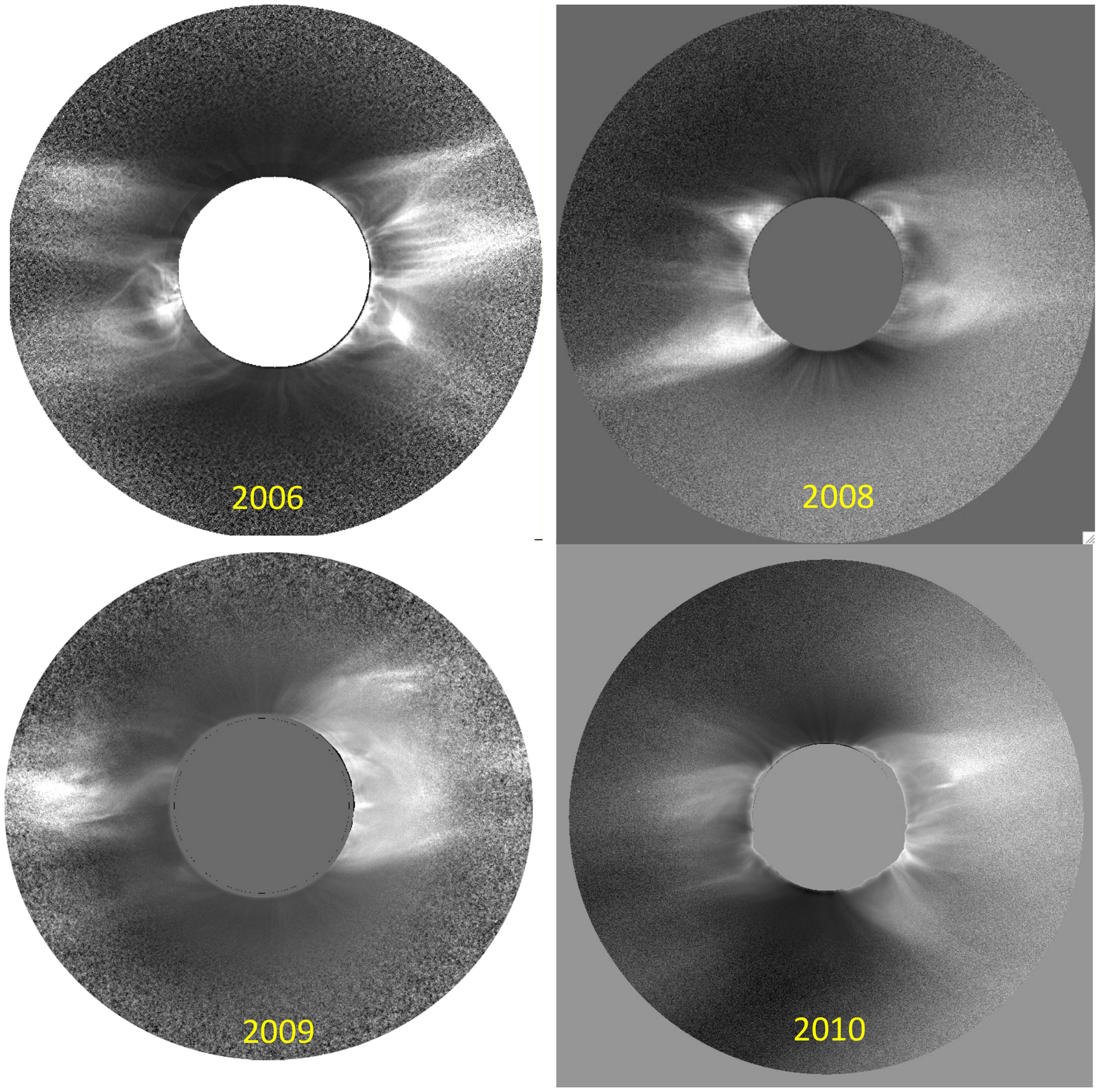}
\includegraphics[width=1.6in]{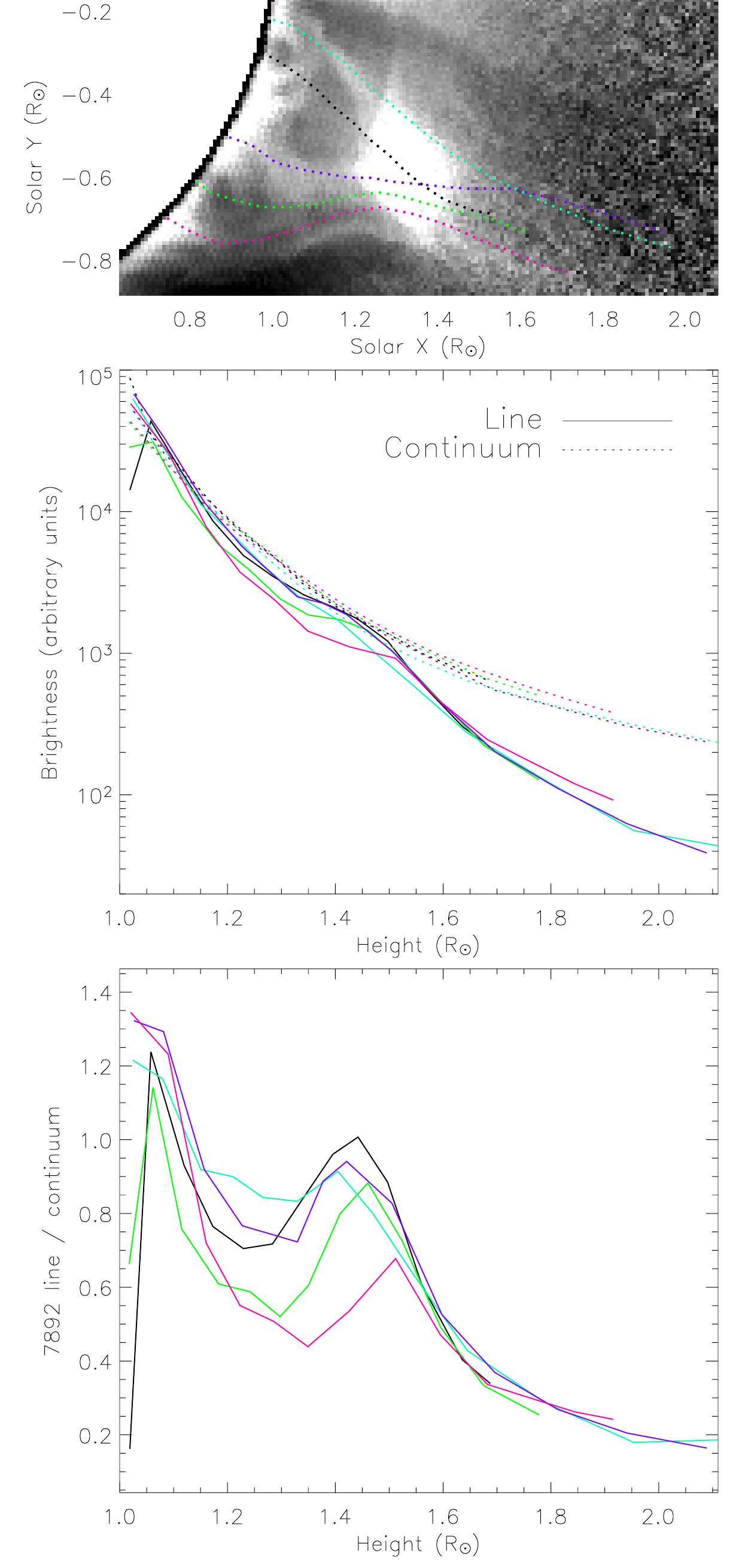}}
\caption{ \small Left four panels: Fe XI 789.2 nm images of the corona taken during the eclipses of 2006, 2008, 2009 and 2010, and processed with NRGF (Morgan et al. 2006). This technique compensates for the radial falloff of intensity with distance without introducing any spurious artifacts in the images. Right panel: Trapping of ions in select magnetic structures as evidenced by the profiles of the Fe XI 789.2 nm line intensity (middle) and intensity ratio of Fe XI 789.2 nm and neighboring continuum at 787 nm (bottom) for the southwest streamers in the 2006 observations. Each colored curve is derived from a different coronal structure as indicated by the traces in the top panel. The transition to a predominantly resonantly scattered emission becomes evident when the slope of the ratio changes from a rapid falloff to an almost constant value as a function of radial distance. }
\label{fexi}
\end{figure}

\begin{figure}[ht]
\centerline{
\includegraphics[scale=0.5]{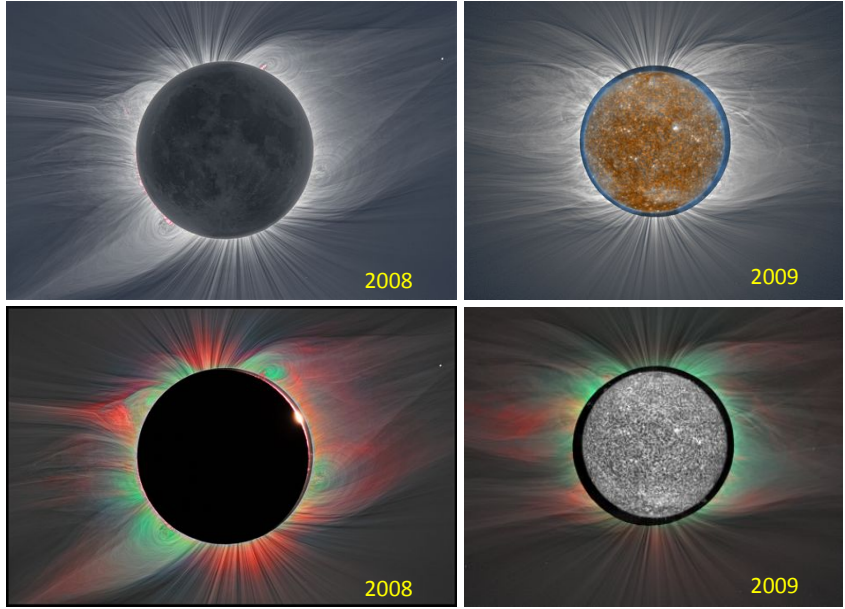}}
\caption{\small White light (top panels) and overlays of Fe XI 789.2 nm (red), Fe XIV 530.3. nm (green), and white light (lower panels) for the 2008 (left) and 2009 (right) eclipses.  He II 304 SoHO/EIT are overlaid on the solar disk for the 2009 eclipse. Images have been processed using the technique developed by Druckm\"uller (2009). }
\label{eclipses0809}
\end{figure}

The realization that imaging the corona in the visible wavelength range provides unique diagnostic capabilities, compared to their EUV counterpart, emerged from the 2006 eclipse observations of the Fe XI 7892 \AA\ line. In the composite  of four eclipse observations of this spectral line (Figure \ref{fexi}), one outstanding feature is the unexpected extent of the Fe XI emission, out to at least 3 \Rs\ in streamers.  The other striking feature is the appearance of regions with localized intensity enhancements at distances ranging from 1.2 to 1.5 \Rs.   These enhancements are often visible in some wavelengths and not others, as they are a function of the underlying plasma temperature. Curiously, there are also features in Fe XI that are much more striking than in their white light counterpart, as seen in Figure \ref{eclipses0809} (see Habbal et al. 2007, 2010a for details). Also striking are the twists in the large scale coronal structures, which are often associated with the magnetic field and density. 

The observed radial extent of the emission from the coronal lines is a consequence of the dominance of radiative relative to collisional excitation in the formation of these lines, within a few tenths of a solar radius above the limb. Beyond that distance the intensity of the emission decreases with radial distance in proportion to the ion density. This is most pronounced for the Fe XI line (see Figure 7 in Habbal et al. 2007a) in comparison with the more 'standard' red (Fe X 637.4 nm) and green (Fe XIV 530.3 nm) lines. The extreme ultraviolet lines observed by EIT/SOHO, EUVI/STEREO and SDO, on the other hand, are dominated by collisional excitation. Their intensity thus falls off very quickly with radial distance as the electron density squared.  While the maximum extent of the detectable UV emission is 1 \Rs\ above the limb, emission from the visible and near infrared Fe lines extends beyond 2 \Rs, and emission lines of other elements can be observed as well. 

The localized intensity enhancements, first observed in Fe XI, are a signature of increases in the Fe$^{+10}$ density relative to that of electrons (see details in Figure \ref{fexi}, right panel). Model studies show that the localized ion abundance enhancements have implications for the distribution of heating in the corona, which also depends on the underlying magnetic geometries (Lie-Svendsen and Esser 2005).

\begin{figure}[ht]
\centerline{
  \includegraphics[width=2.in]{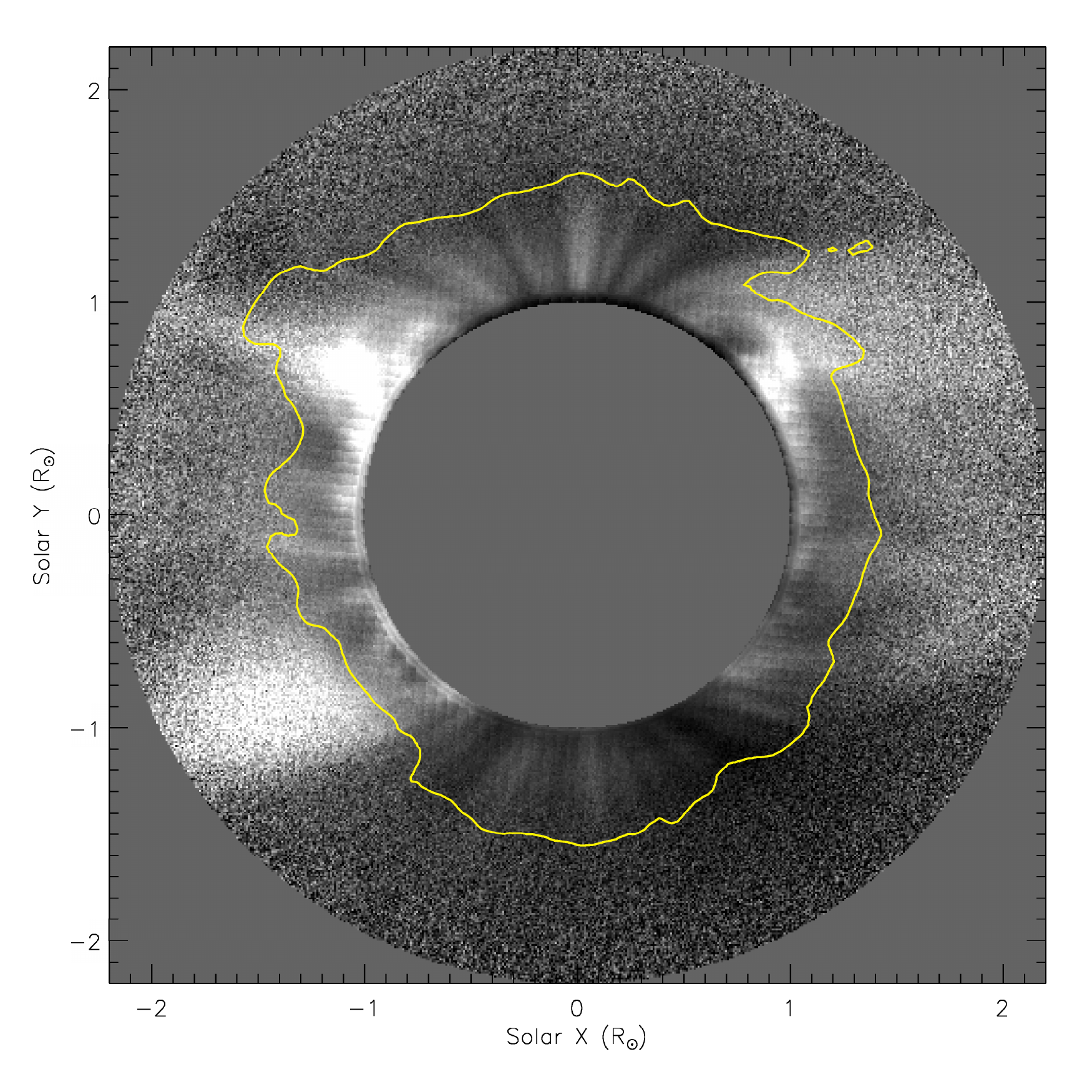}
  \includegraphics[width=2.in]{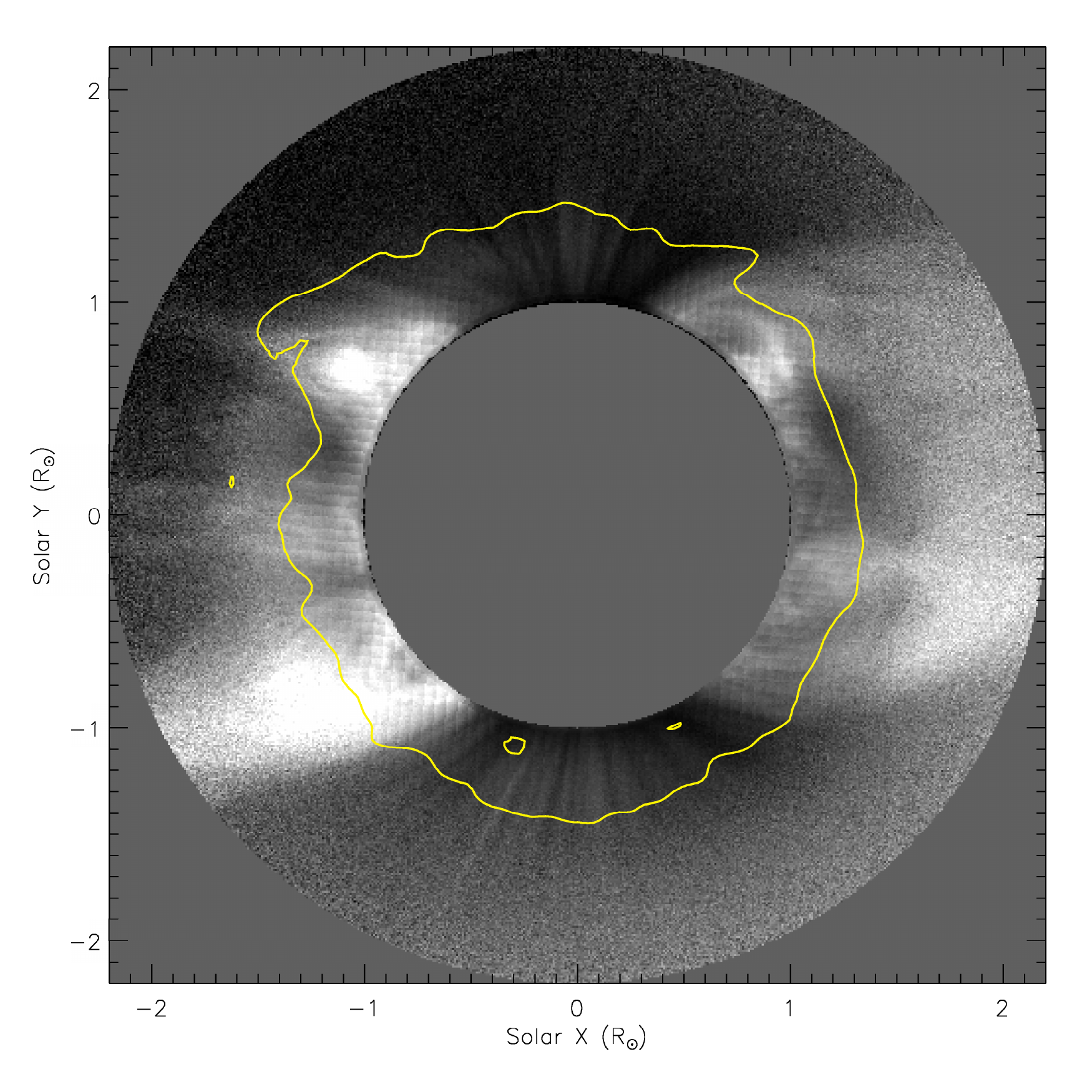}}
\centerline{
  \includegraphics[width=2.in]{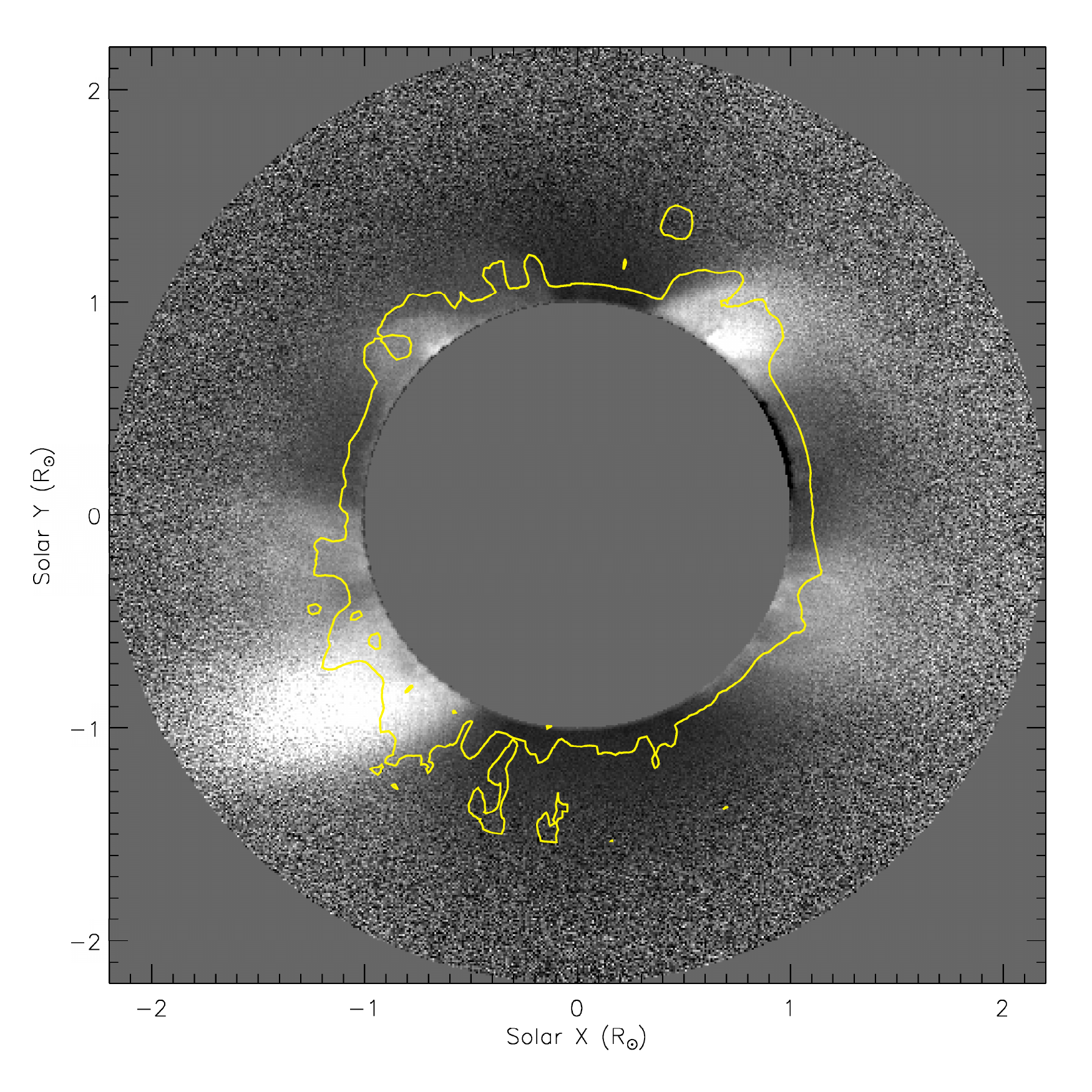}
  \includegraphics[width=2.in]{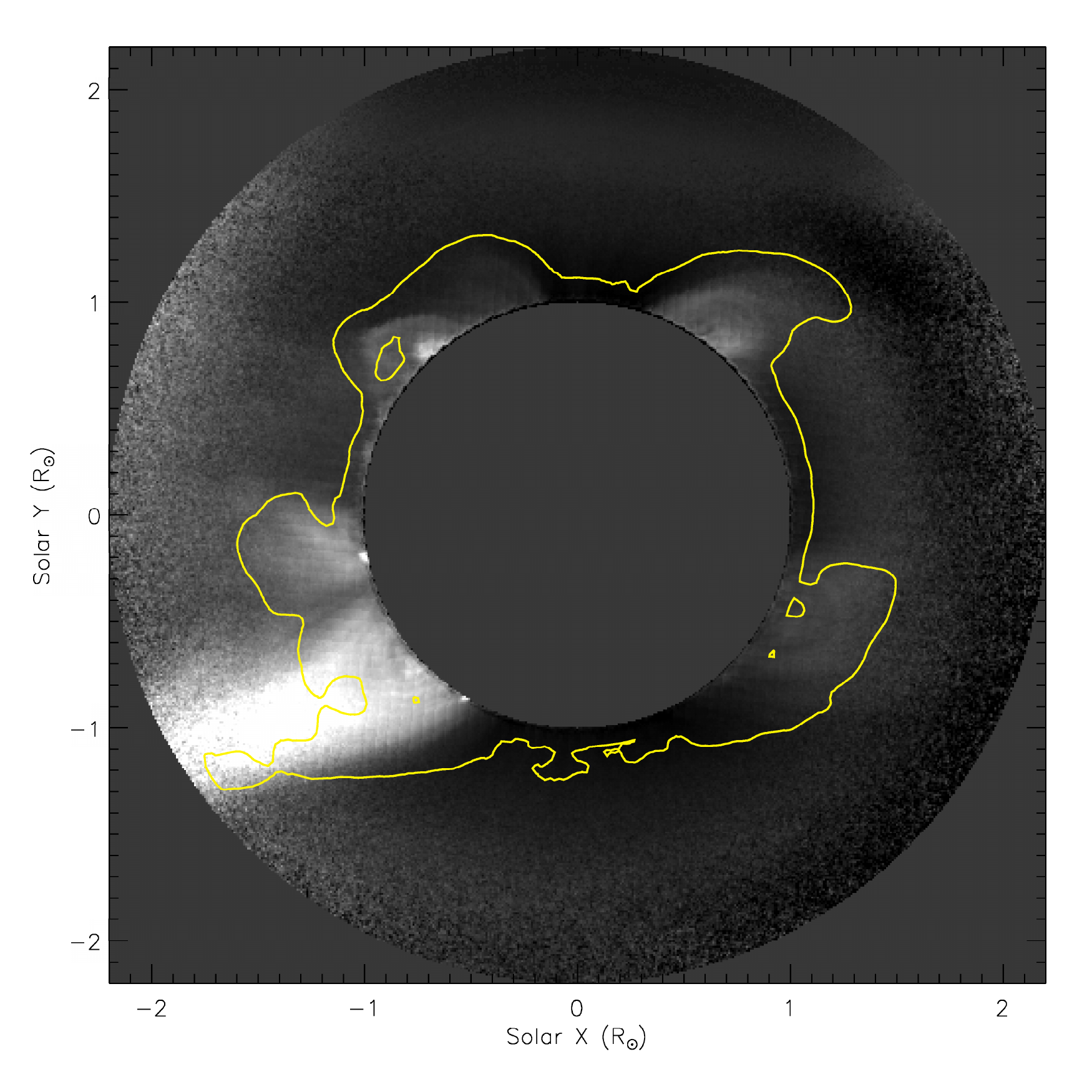}}
\caption{\small Contours of $R_t$ overlayed on the Fe X (top left), Fe XI (top right), Fe XIII (bottom left) and Fe XIV (bottom right) 2008 eclipse NRGF-processed images. (From Habbal et al. 2010a).}
\label{rtcontour}
\end{figure}

Comparison with continuum observations yields the distance at which the emission transitions from being dominated by collisions to one dominated by resonant scattering, as shown in the bottom right panel of Figure \ref{fexi}. Such a transition thus corresponds to the transition from a collision-dominated to a collisionless plasma. As shown in Figure \ref{rtcontour}, such a determination can be made for each of the spectral lines observed. Since this transition does not occur at the same place for all the lines, implies a dependence on the density and temperature of the underlying magnetic structure (Habbal et al. 2010a). The locus of the transition, $R_t$, between a collision-dominated regime to a collisionless one for a given emission line,  ranging between 1.1 and 2 \Rs, also determines the locus of the plasma's departure from ionization equilibrium (Habbal et al. 2010c).  

Another unique feature of these two-dimensional multi-wavelength observations is the inference of the two-dimensional distribution of the electron temperature in the corona. As seen in Figure \ref{eclipses0809}, there is a clear demarcation between emission in Fe X/FeXI (covering a temperature range of 0.8 to $1.2 \times 10^6$ K) and Fe XIII/FeXIV (characteristic of 1.6 - 2.2 $\times 10^6$ K plasmas) for most coronal structures. There is also a distinct dominance of cooler Fe X and Fe XI emission in polar regions. The presence of emission from all spectral lines, such as in some streamers in both 2008 and 2009 observations, is a reflection of a multi-temperature plasma there. The two-dimensional multi-wavelength composites can thus be regarded as maps of the coronal electron temperature placed in the context of density/magnetic structures. The 2006 - 2009 observations indicate that the expanding corona is dominated by cooler electron temperatures averaging $10^6$ K (Habbal et al. 2010a).

\begin{figure}[ht]
\centerline{
\includegraphics[scale=0.45]{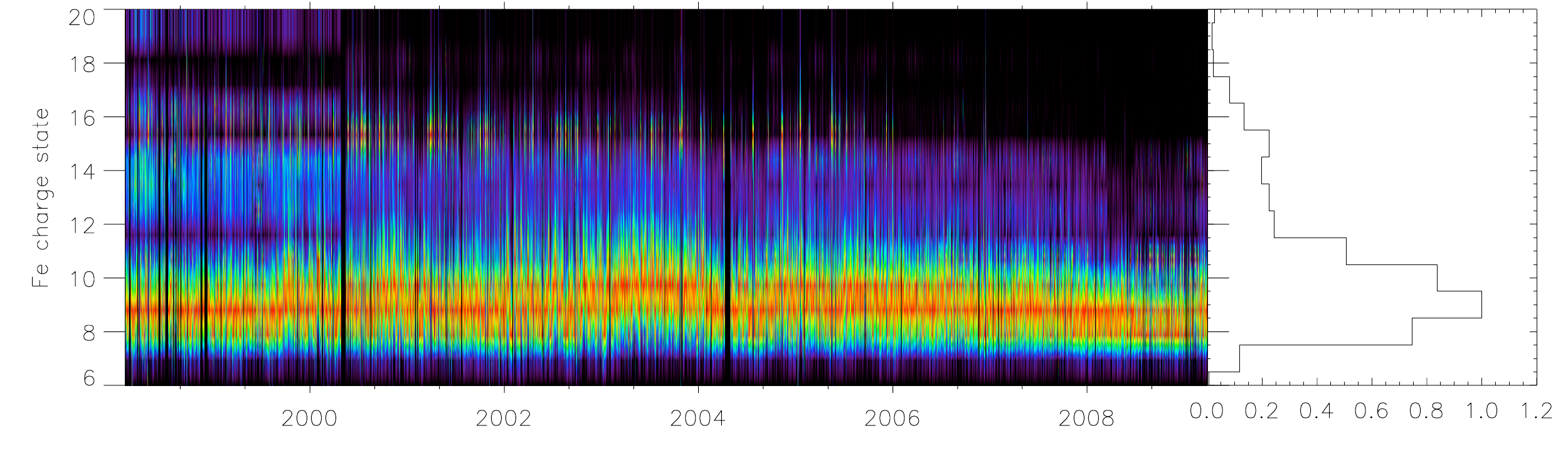}}
\centerline{
  \includegraphics[scale=0.3]{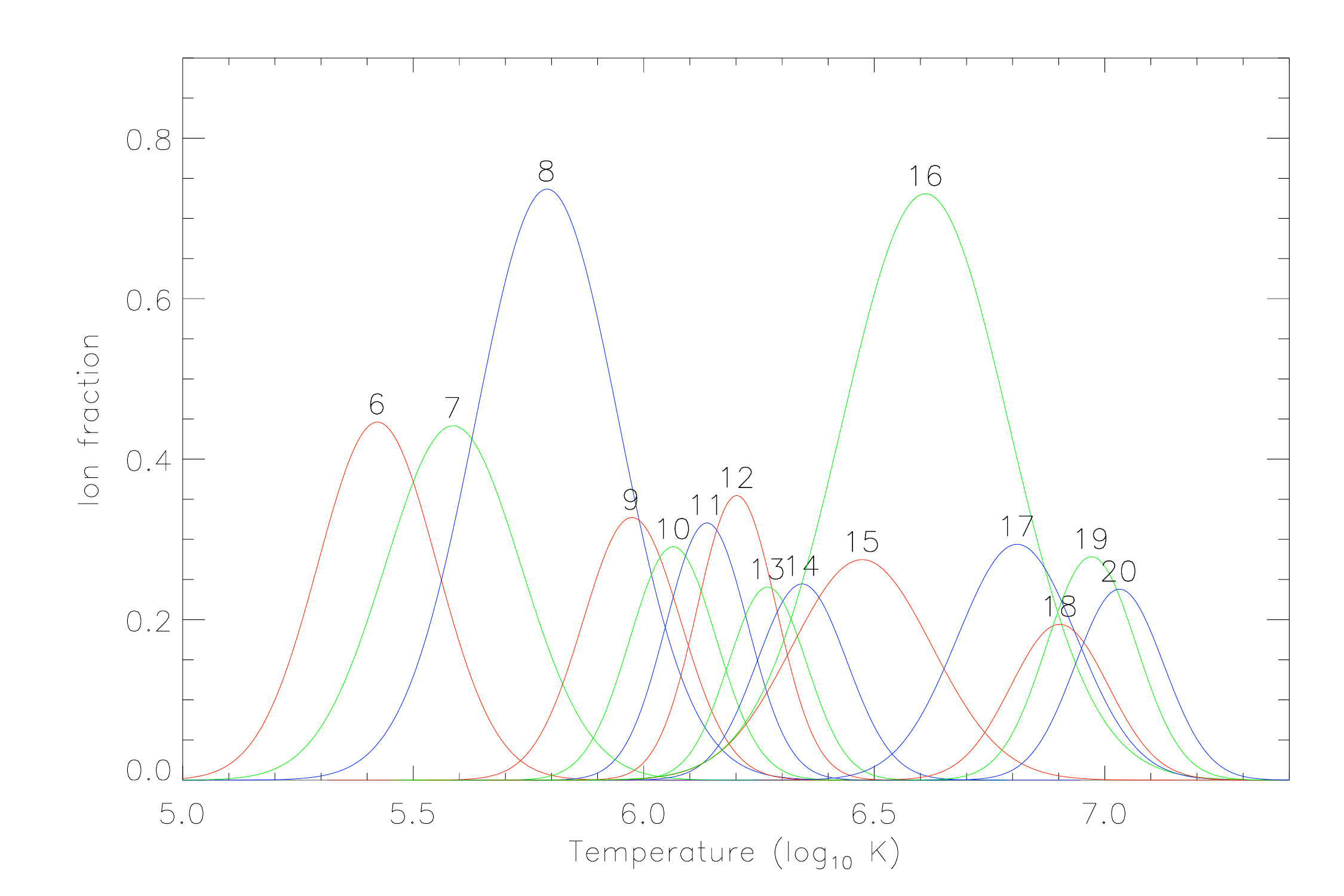}
  \includegraphics[scale=0.3]{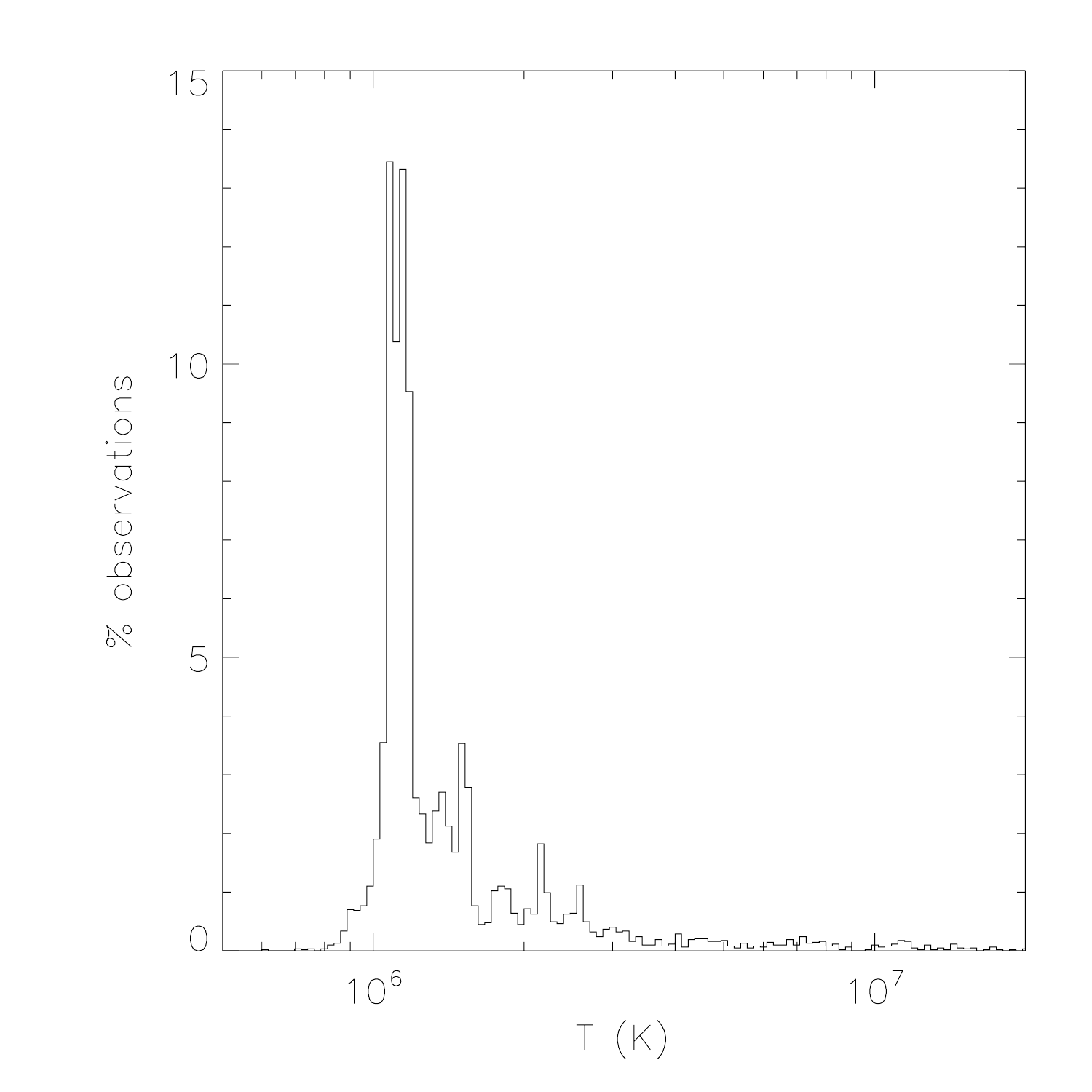}  }
\caption{\small Top: 12 hour averages of the SWICS/{\it ACE} Fe$^{n+}$ (with n = 6 to 20) charge state measurements as a function of time, from 1998 to 2009 (left), and their histogram averaged over the entire time interval, and normalized to unity (right). Lower left: Calculated Fe$^{n+}$ ion fraction, versus temperature for the measured charge states, using data published by Mazzotta et al. (1998) cited in the CHIANTI database. Lower right: Calculated distribution of coronal electron temperatures that best reproduces the SWICS/{\it ACE} Fe charge state data above. The vertical axis is the percent of observations in each temperature bin.  (From Habbal et al. 2010b).}
\label{ace1}
\end{figure}

The characteristic distance $R_t$ where the transition from a collisional to a collisionless plasma occurs for each spectral line also represents the distance beyond which the emission reflects the distribution of its corresponding charge state. In the examples of 2006, 2008 and 2009, the emission was dominated by Fe XI implying a dominance of Fe$^{+10}$ charge states in the expanding corona. The only charge state measurements available for comparison are those from interplanetary space, which show a distinct peak in the ion fraction of iron at  Fe$^{+10}$  in both the fast and slow wind during the periods covered by the eclipse observations. As shown in more detail in Figure \ref{ace1}, the dominance of an electron temperature of $10^6$ K and the Fe$^{+10}$ charge state in the corona match remarkably well the dominance of Fe$^{+10}$ charge state measured in interplanetary space. This correspondence establishes an unambiguous link between conditions in the inner corona and interplanetary space (see Habbal et al. 2010b).

The multi-wavelength eclipse observations also revealed the role of prominences in defining large scale coronal structures and their underlying thermodynamic properties. As evident in the examples of Figure \ref{eclipses0809}, and described in detail in Habbal et al. (2010c), prominences are enshrouded in hot plasmas of twisted magnetic structures (formally referred to as 'coronal cavities') which form the bases (or bulges) of streamers.  The observed twisted helical structures within the bulges of streamers provide the most direct evidence for the emergence of helicity with prominences that is not limited to the prominences themselves but extends to their immediate surroundings. The fact that twisted structures surrounding cool prominence material are tenuous and hot, often exceeding $2 \times 10^6$ K, should have significant implications for coronal heating mechanisms in the neighborhood of magnetic polarity reversal regions.   

 \begin{figure}[ht]
\centerline{
\includegraphics[width=4.in]{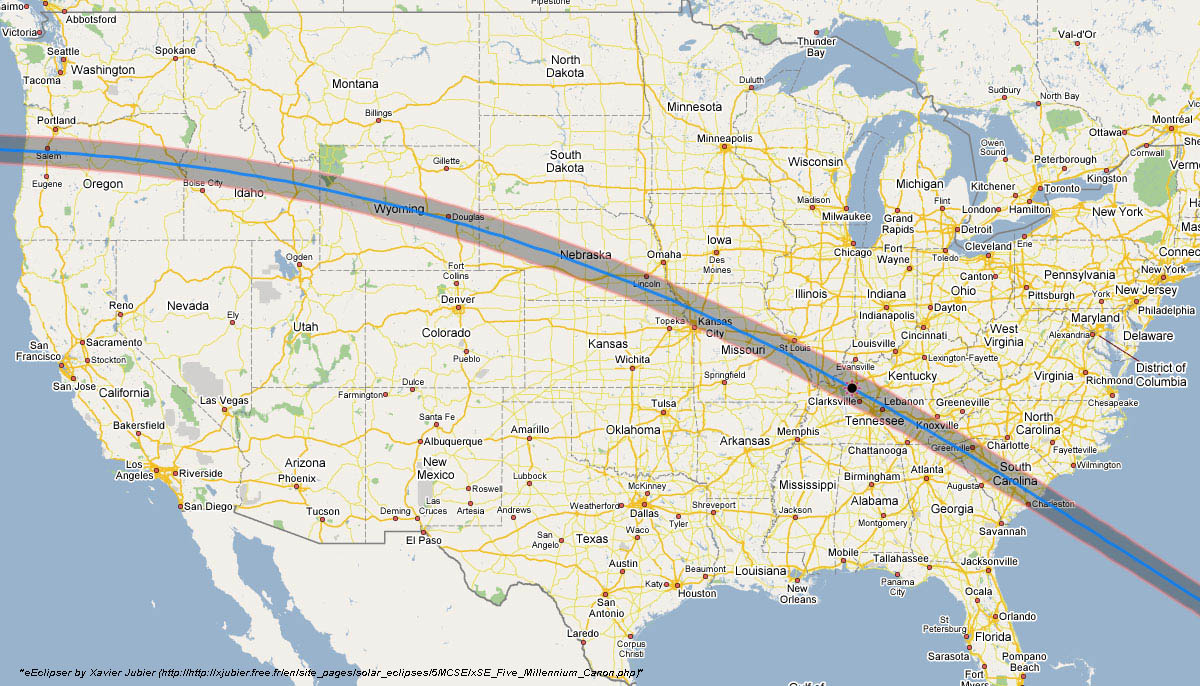}}
\caption{\small Path of the 21 August 2017 eclipse across the US, with a total duration of 1h 33min 16.8s. }
\label{eclipsepath2017}
\end{figure}

\section*{The Case for Support for Total Solar Eclipse Observations in the Next Decade}

As described above, the multi-wavelength eclipse observations in the visible and near IR  lines of Fe taken between 2006 and 2010 led to a number of key insights into the magnetic and density structures of the corona, starting from the solar surface out to a few solar radii. 
The key property that enabled these discoveries is the radiative component in the emission process for these spectral lines, which dominates over the collisional component as the view moves away from the solar limb. The other important property is the well-defined temperature diagnostic offered by these coronal emission lines, as thin-film interference filters for the visible and near infrared are available with bandpasses even narrower than the Doppler-broadened coronal line widths. Consequently, spectral lines are well-separated and line profiles can even be explored. Such is not the case for EUV lines currently used in space-based observatories for temperature and density diagnostics of the corona, as multilayer coatings for EUV imaging provide wide bandpasses with multiple spectral lines in them. 
The recent eclipse observations provided new insight into the role of selective heating of minor ions in the inner corona, and how it leads to localized ion density enhancements, and is linked to the underlying magnetic structures. They established the locus of the transition from a collision-dominated to a collisionless regime. They also provided the first two-dimensional maps of the coronal electron temperature and coronal Fe charge states. 

In addition to these examples, two very important diagnostic capabilities of the visible and near IR spectral ranges have to be kept in mind, namely the role of spectroscopy and polarimetry. Despite the proliferation of space-based observations of the solar corona starting with the OSO missions and Skylab in the late 60's and early 70's, spectra of the corona in the visible wavelength range remain rare, with the most comprehensive ones dating back to the 1960's. However, spectroscopy without contextual images falls short in yielding its full potential, as is evident from the different published accounts of some spectral lines recorded at some eclipses but not at others (e.g., Jefferies et al. 1971). Indeed, it is only when the underlying magnetic and density structures are identified in the corona, that one can make sense of the corresponding spectrum, i.e. the presence or absence of a given spectral line can be accounted for when the context of the spectrum is known, such as H$\alpha$ being present in prominences, but not in other regions of the corona.  Polarimetric observations in select spectral lines (Judge 1998) offer a powerful tool for coronal magnetometry (Lin et al. 2000). Although they are difficult measurements, primarily because of the stringent requirements on observing conditions, they are the only tools currently available for inferring the direction and strength of the coronal magnetic field. Polarization measurements carried out with different spectral lines will enable the identification of the coupled temperature-magnetic structures. Furthermore, the coupled polarization and imaging measurements can yield the locus of the transition from closed to open magnetic structures. Knowledge of this transition is essential for model studies of coronal heating and solar wind acceleration.  

Unfortunately, no space-based coronagraph at present has any capability for narrowband imaging, spectroscopy or polarimetry in the visible and near IR wavelength range.  Furthermore, even with technological advances in formation flying and/or deployable boom technology that would support a larger occulter at a greater distance, the level of stray light suppression would not match that available from occultation by the moon. At ground-based coronagraphs there is the additional issue that sky conditions are no match to the reduction in sky brightness during eclipses.  A total solar eclipse provides a unique opportunity to observe the extent of the emission from the solar limb out to a few solar radii, the key to exploring the most critical region of space involving the evolution of the coronal magnetic field and the acceleration of the solar wind. 

A space-based alternative to coronagraphs that would overcome the stray light issues would be an observatory in lunar orbit, providing a daily sequence of eclipse periods that would allow for coronal observations by a small instrument which could be part of an observatory for lunar and heliophysical science.  

Given the limitations of our current space- and ground-based observatories, it seems that eclipse observations offer unique opportunities for further exploring the physics of the corona in the least ambiguous manner. The impetus for these observations can also drive the development for new detector technology for observations in the near IR. Furthermore, there is no doubt that such scientific endeavors cannot be carried out without the underlying support provided by the NASA eclipse bulletins that continue to be prepared by F. Espenak and J. Andersen, which needs to be secured. The most desirable eclipse for organizing serious and coordinated observations is the 21 August 2017 eclipse which will cross the United States (see Figure \ref{eclipsepath2017}). The total duration of this eclipse across the US of over an hour provides a further unprecedented opportunity for detailed observations of the spatial scales of coronal structures, and their evolution and variation in time.



\begin{thebibliography}{}



\bibitem[Druckm{\"u}ller(2009)]{druckmuller2009} Druckm{\"u}ller, M. 2009, \apj, 706, 1605

\bibitem[Druckm{\"u}ller et al.(2006)]{druckmuller2006} Druckm{\"u}ller, M., Ru{\v s}in, V., 
\& Minarovjech, M.\ 2006, Contributions of the Astronomical Observatory Skalnate Pleso, 36, 131 







\bibitem[Habbal et al.(2007a)]{hab2007a} Habbal, S.~R., Morgan, H., Johnson, J., Arndt, M.~B., Daw, A., Jaeggli, S., Kuhn, J., \& Mickey, D.\ 2007a, \apj, 663, 598 

\bibitem[Habbal et al.(2007b)]{hab2007b} Habbal, S.~R., Morgan, H., Johnson, J., Arndt, M.~B., Daw, A., Jaeggli, S., Kuhn, J., \& Mickey, D.\ 2007b, \apj, 670, 1521 

\bibitem[Habbal et al.(2010a)]{hab2010a} Habbal, S.~R., et al. 2010a, \apj, 708, 1650

\bibitem[Habbal et al.(2010b)]{hab2010b} Habbal, S.~R., Morgan, H., Druckm\"uller, M., \& Ding, A. 2010b, \apj, 711, L75

\bibitem[Habbal et al.(2010c)]{hab2010c} Habbal, S.~R., et al. 2010c, \apj, 719, 1362


\bibitem[Jefferies(1971)]{jeff1971} Jefferies, J. T., Orrall, F. Q., \& Zirker, J. B. 1971, \solphys, 16, 103

\bibitem[Judge(1998)]{judge1998} Judge, P. G., \apj, 500, 1009-1022, 1998




\bibitem[Lie-Svendsen(2005)]{lie-sven2005} Lie-Svendsen, O \& Esser, R. 2005, \apj, 618, 1057

\bibitem[Lin et al. (2000)]{lin2000} Lin, H., Penn, M. J., and Tomczyk, S. 2000, \apj., 541, L83



 




\bibitem[Morgan et al.(2006)]{morgan2006}
Morgan, H., Habbal, S.~R., \& Woo, R. 2006, \solphys, 236, 263




\bibitem[Pasachoff et al.(2009)]{pas2009} Pasachoff, J.~M., Ru{\v s}in, V., Druckm\"uller, M., Aniol, P., Saniga, M., 
\& Minarovjech, M.\ 2009, \apj, 702, 1297

 


\end{thebibliography}
\end{document}